\documentclass[aps,prl,twocolumn,groupedaddress,showpacs,superscriptaddress]{revtex4-1}
\usepackage{graphicx}
\usepackage{amssymb}
\usepackage{amsmath}

\begin{document}


\title{Accurate Measurement in the Field of the Earth of the General--Relativistic Precession of the LAGEOS II Pericenter and New Constraints on Non--Newtonian Gravity}


\author{David M. Lucchesi}
\affiliation{Istituto di Fisica dello Spazio Interplanetario,
Istituto Nazionale di Astrofisica, IFSI/INAF,\\
        Via del Fosso del Cavaliere 100, 00133 Roma, Italy}
\affiliation{Istituto di Scienza e Tecnologie dell'Informazione, Consiglio Nazionale delle Ricerche, ISTI/CNR, \\
        Via G. Moruzzi 1, 56124 Pisa, Italy}
\author{Roberto Peron}
\affiliation{Istituto di Fisica dello Spazio Interplanetario,
Istituto Nazionale di Astrofisica, IFSI/INAF,\\
        Via del Fosso del Cavaliere 100, 00133 Roma, Italy}



\begin{abstract}
The pericenter shift of a binary system represents a suitable
observable to test for possible deviations from the Newtonian
inverse--square law in favor of new weak interactions
between macroscopic objects. We analyzed 13 years of tracking data of the LAGEOS
satellites with GEODYN II software but with no models for
general relativity. From the fit of LAGEOS II pericenter residuals
we have been able to obtain a 99.8\% agreement with the predictions
of Einstein's theory. This result may be considered as a
99.8\% measurement in the field of the Earth of the combination of
the \(\gamma\) and \(\beta\) parameters of general relativity, and it may be
used to constrain possible deviations
from the inverse--square law in favor of new weak
interactions parametrized by a Yukawa--like potential
with strength \(\alpha\) and range \(\lambda\). We obtained
\(|\alpha| \lesssim 1 \cdot 10^{-11}\),
a huge improvement at a range of about 1 Earth radius.
\end{abstract}


\maketitle

Tests for Newtonian gravity and for a possible violation of the weak
equivalence principle are strongly related and represent a powerful
approach in order to validate Einstein's theory of general relativity
(GR) with respect to alternative theories of gravity and to tune,
from the experimental point of view, gravity itself into the realm
of quantum physics. Moreover, new long range interactions (NLRIs) may
be thought of as the residual of a cosmological primordial scalar
field related with the inflationary stage (dilaton scenario)
\cite{2002PhRvD..66d6007D}. Twenty--four years ago, the possibility
of a fifth force of nature prompted new experimental investigation
of possible deviations from the gravitational inverse--square law
\cite{1986PhRvL..56....3F}. In fact, the deviations from the usual
\(1/r\) law for the gravitational potential would lead to new weak
interactions between macroscopic objects.

Interestingly, these supplementary interactions may
be either consistent with Einstein's equivalence principle or not. In
this second case, nonmetric phenomena will be produced with tiny,
but significant, consequences in the gravitational experiments
\cite{1998Nordtvedt}. The feature of such interactions, which
are predicted by several theories, is to produce deviations for
separations of masses ranging through several orders of magnitude,
starting from the submillimeter level up to the astronomical
scale. Among the various techniques useful for the search of this
additional physics to the various scales, the accurate measurement
of the pericenter shift of binary systems may be used to test for a NLRI with a
characteristic range comparable with the system semimajor axis
\cite{2000PhRvD..61l2001N}.

These very weak NLRI are usually described by means of a
Yukawa--like potential with strength \(\alpha\) and range
\(\lambda\) and transmitted by a field of very small mass
\(\mu=\hbar/\lambda c\). If \(G_{\infty}\) represents the
gravitational constant, \(M_{\oplus}\) and \(m_s\) the mass of the
primary body and of the satellite, \(r\) their
separation, \(c\) the speed of light and \(\hbar\) the reduced
Planck constant, we can write:

\begin{equation}
\label{yukawa} V_{Yuk}=-\alpha{G_{\infty}M_{\oplus} \over
{r}}e^{-r/\lambda}, \quad \alpha={1 \over
{G_{\infty}}}\left({K_{\oplus} \over {M_{\oplus}}}{K_s \over
{m_s}}\right),
\end{equation}

where the strength \(\alpha\) depends both on the mass--energy
content of the sources and on their coupling strengths,
\(K_{\oplus}\) and \(K_s\), respectively.

In the weak field and slow motion limit (WFSML) of
GR, Einstein's equations reduce to a form quite similar to those
of electromagnetism \footnote{Masses produce a gravitoelectric field,
analogous to the electric field produced by charges, while mass
currents produce a gravitomagnetic field, analogous to the magnetic
field produced by electric currents.}.
In Einstein's geometrodynamics and in the frame of a relativistic
3--body problem, where the two primaries are the Sun and the Earth
and the test particle is represented by a satellite orbiting the
Earth, the main precessions to which the satellite orbit, as a sort
of enormous gyroscope, is subject to are commonly known in the
literature as: i) Einstein \cite{1916AnP...354..769E}, ii) de Sitter
\cite{1916MNRAS..77..155D}, and iii) Lense--Thirring (LT)
\cite{1918Lense-Thirring,*1984GReGr..16..711M} precessions.
These precessions may be explained in terms of
the effects, on the orbital plane of the satellite, produced by the
gravitoelectric and gravitomagnetic fields of the Earth, points i)
and iii), and by the effects arising from the coupling between the
Earth's motion with the background field of the Sun, point ii). While
Einstein's precession is a spin--independent secular effect, the
other two precessions are usually interpreted as spin--orbit
effects, in particular, as frame--dragging effects, but with some
differences: the de Sitter precession is frame--dependent, while the
LT one is intrinsically related with the spin of the primary mass,
i.e., with its rotation. Therefore, this precession must be related
with \emph{intrinsic} gravitomagnetism. We refer to Ciufolini and
Wheeler \cite{1995grin.book.....C} and to Ciufolini
\cite{2010NewA...15..332C} for a deeper insight.

This Letter is devoted to showing some of our recent results on the
first simultaneous measurements of the cited relativistic precessions
in the field of the Earth using the two LAGEOS (LAser GEOdynamics
Satellite) satellites. This work is new with respect to previous
ones because we measure all the relativistic secular effects at one
time. In particular, we focus on the satellites' pericenter secular
advances \cite{1977CeMec..15...21R}, to which several non--Newtonian
theories of gravity are sensitive. We analyzed 13 years of Satellite
Laser Ranging (SLR) data of the two LAGEOS satellites using the
NASA/GSFC software GEODYN II \cite{1998pavlis}. This software is
dedicated to satellite orbit determination and prediction, geodetic
parameters estimation, tracking instruments calibration, and many
other applications in the field of space geodesy. The key
ingredients of our measurement are (i) a consistent statement of the
theory to be tested, (ii) the availability of a good test mass with
related high--quality tracking data and (iii) a modelization set for
test mass dynamics and tracking.

The first ingredient is far from being trivial: the relativistic
equations of motion can be formulated in principle (due to general
covariance) in whatsoever coordinate system; however, see Ashby and
Bertotti (AB) \cite{1984PhRvL..52..485A}, a suitable choice of this
system makes its physical interpretation clearer and simplifies its
formulation. In their generalized local inertial frame the main
contribution to the test mass dynamics comes from the central body,
while third--body effects show up only through tidal
terms. The GR acceleration model included in GEODYN II follows the
results of Huang \emph{et al.} \cite{1990CeMDA..48..167H}, and represents a
generalization of the AB model. The main feature of this model is
that their noninertial geocentric frame retains all the merits of
the inertial geocentric frame of AB, but it does not rotate with
respect to the barycentric reference frame.

The second ingredient is given by the laser ranging data of LAGEOS
satellites. The two are almost twins \cite{2007Andres}. LAGEOS,
launched by NASA (1976), and LAGEOS II, launched by NASA/ASI (1992),
have been designed spherical in shape, with high density and small
area--to--mass ratio in order to minimize the effects of the subtle
and complex nongravitational perturbations
\cite{1987nongrav.book.....M}. Their radius is just 30 cm and their
mass about 407 kg \footnote{LAGEOS has an almost circular orbit,
with an eccentricity \(e_{I} \simeq 0.004\), a semimajor axis
\(a_{I} \simeq 12,270\)~km and an inclination over the Earth's
equator \(i_{I} \simeq 109.8^{\circ}\). LAGEOS II corresponding
elements are: \(e_{II} \simeq 0.014\), \(a_{II} \simeq 12,162\)~km
and \(i_{II} \simeq 52.66^{\circ}\).}. Their aluminum surface is
covered with 426 cube--corner retro--reflectors for laser ranging
from dedicated ground stations. The precision of the measurements is
mainly related with the pulse width, which is usually
\(\approx\)~\(1\cdot10^{-10}\)~s down to \(3\cdot10^{-11}\)~s for
the best laser ranging stations.
The SLR data are available through
the International Laser Ranging Service (ILRS)
\cite{2002AdSpR..30..135P} in the form of normal points, with a
root--mean--square (rms) down to a few mm, that corresponds to an
accuracy in the orbit reconstruction at a few cm level, when using
the best dynamical models. In our preliminary analyses we have been
able to fit the orbit of the satellites at a 1--2~cm (rms) level in range.

Regarding the third ingredient, the models included in
GEODYN II are devoted to describe satellite dynamics, measurement
procedure, and reference frame transformations; they include
\cite{2004IERS-Conv-2003}: (i) the geopotential (static and dynamic),
(ii) lunisolar and planetary perturbations, (iii) solar radiation
pressure and Earth's albedo, (iv) Rubincam and Yarkovsky--Schach
effects (which need the satellite spin--axis coordinates), (v) SLR
stations coordinates, (vi) ocean loading, (vii) Earth Orientation
Parameters and (viii) measurement procedure. Usually, the models
implemented in the code include the GR corrections
in the parametrized post--Newtonian (PPN) formalism
\cite{1968PhRv..169.1017N,*1971ApJ...163..611W}.

In the analysis we performed, in order to solve for the relativistic
secular precessions, we did not include in our setup such
corrections. Moreover, we did not estimate any empirical
accelerations as well as the satellites' radiation coefficient
\(C_R\), polar motion and universal time UT1 corrections, in such a
way to avoid any possible absorption of physical effects. Finally,
in order to avoid the problems related with the spin modeling, also
the Rubincam and Yarkovsky--Schach (YS) effects have not been included in
our analysis. Concerning the estimated parameters, besides the satellite
state--vector, for each 15--day arc we estimated only measurement biases.
The long--arc analysis of the orbits of geodetic satellites is a
useful way to extract relevant information (i.e., model parameters)
concerning the Earth's structure \footnote{It is necessary to decompose the
long arc in a number of shorter arcs (15 days in our analysis), not
causally connected, and solve for the satellite state vector for each arc together with a set of
parameters in order to absorb unmodeled or poorly modeled
perturbations over the arc.}. The physical information is concentrated
in the satellite orbital residuals, that must be extracted from the
orbital elements determined during the fit
\cite{2006P&SS...54..581L}.
The software fits the tracking data with all its models
minimizing the difference between the observed data and the computed
ones, using a differential correction procedure.
The final residuals are a measure of all the unmodeled
effects, such as the GR ones, as well as of the poorly modeled and
mismodeled effects and the noise in the tracking data.
The relativistic precessions are effective in the orbital elements
that define the orbit orientation in space, i.e., the orbit Euler
angles with respect to Earth's equatorial plane. These are the
longitude \(\Omega\) of the orbit ascending node, the argument of
pericenter \(\omega\), and the orbit inclination \(i\). These
elements are not equally sensitive, in their secular and periodic
effects, to the relativistic precessions. Anyway, the argument of
pericenter \(\omega\) is sensitive to all of them, in particular to
their secular effects.

In our analysis we determined the orbital elements of the two LAGEOS
satellites and then we computed the residuals with the method
explained in Ref. \cite{2006P&SS...54..581L}. This is the method
developed by one of us in 1996, and that has been always used in the
LT effect measurements performed so far
\cite{1996NCimA.109..575C,*1997EL.....39..359C,*1998Sci...279.2100C,*2004Natur.431..958C,*2006NewA...11..527C}.
This point represents a crucial aspect in this kind of analysis,
because we need a reliable way to obtain the residuals in the
orbital elements which retain the original concept of
\emph{observed} - \emph{computed} quantity, which usually refers to
the tracking observable, the range for SLR data. With
regard to the background gravity field, in our setup we included two
different models: (1) EGM96 and (2) EIGEN-GRACE02S. The gravity
field plays a very significant role in this kind of measurement. The
uncertainties in its harmonic coefficients, especially in the even
zonal ones, are the major source of systematic effects, as we know
very well in the case of the previous measurements of the LT effect.
The EGM96 model \cite{1998Lemoine}, the current conventional model
recommended by the International Earth Rotation Service (IERS) \cite{2004IERS-Conv-2003},
is a multisatellite model derived over a time span of several years. The advantage of a
multisatellite model resides in the different orbital
characteristics of the satellites, such as different semimajor axis,
eccentricity and inclination \footnote{In this way, the gravity
model has more or less the same sensitivity to the low and the
high degree terms, especially because of the different altitudes of
the satellites. EGM96 is characterized by uncalibrated formal errors
with a relative high correlation between the coefficients.}.
A clear disadvantage is represented by the fact
that the tracking data and the quality of the orbit dynamical
models are not homogeneous for all the satellites. EIGEN-GRACE02S
\cite{2005JGeo...39....1R} has been derived by GRACE mission, and has the
characteristic to improve the
gravity field knowledge with a limited amount of data, in particular
in the medium and long wavelengths of its spectrum.

In the following we focus on the results obtained from the analysis
of the satellite's pericenter shift, in particular, for LAGEOS II.
Indeed, in the case of the pericenter the observable quantity is
\(e\Delta\omega\); i.e., it depends on the satellite eccentricity
\(e\). Because LAGEOS II orbit is more eccentric, LAGEOS II is the
best candidate for an accurate measurement of the total relativistic
precession of the pericenter. In Table \ref{table:1} the results
expected for the relativistic precession rates in the pericenter are shown.
\begin{table}[h]
\begin{center}
\caption{{Rates in mas/yr of the secular relativistic precession on the
argument of pericenter of the two LAGEOS satellites (1 mas/yr = 1
milli--arc--second per year)}.} \label{table:1}
\newcommand{\m}{\hphantom{$-$}}
\newcommand{\cc}[1]{\multicolumn{1}{c}{#1}}
\renewcommand{\tabcolsep}{2pc} 
\renewcommand{\arraystretch}{1.2} 
\begin{tabular}{@{}lcc}
\hline
\text{Rates}  &  \text{LAGEOS II} &  \text{LAGEOS}\\
\hline
\(\Delta\dot{\omega}^E\)     & + 3351.95  & + 3278.77\\
\(\Delta\dot{\omega}^{LT}\)  & - 57.00    & + 32.00\\
\(\Delta\dot{\omega}^{dS}\)  & + 10.69    & - 5.99\\
\hline
\end{tabular}\\[2pt]
\end{center}
\end{table}
By inspection, the total relativistic precession of the LAGEOS II pericenter is
\(\Delta\dot{\omega}_{II}^{rel} \simeq 3305.64\)~mas/yr.

\begin{figure}[h]\abovecaptionskip 1 mm \belowcaptionskip 1 mm
 \includegraphics[width=7cm]{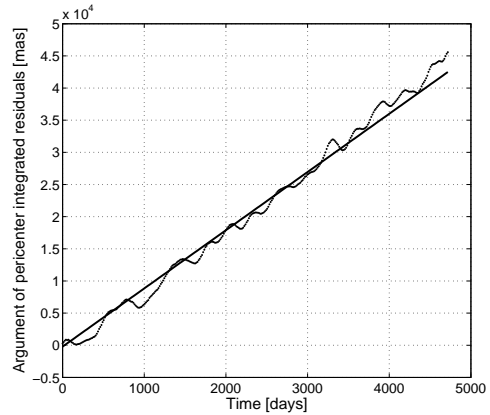}%
 \caption{\label{perigeo}Integrated residuals (dots) of LAGEOS II
 pericenter with the linear trend (continuous line) fitted together with four periodic
 terms with the periods of the main spectral lines of the YS effect.
 The fractional discrepancy between the slope of the linear trend and
 the prediction of general relativity is just \(2.8\cdot10^{-4}\). The starting epoch is MJD 49004.}
 \end{figure}

The result of our 13 year analysis is shown in Figure \ref{perigeo}
for the satellite argument of pericenter advance in the case of the
EIGEN-GRACE02S model. The Fast Fourier Transform (FFT) of the
residuals in the pericenter rate confirms the presence of the
unmodeled YS effect in the integrated residuals of Figure \ref{perigeo}.
We fitted the residuals with a linear trend plus four periodic terms
\footnote{That is, \(\Delta\omega_{fit}=a+bt+\sum_i C_i\sin(2\pi
t/T_i+\phi_i)\), where \(b=\dot{\omega}_{II}^{meas}\) while
\(C_i\), \(T_i\) and \(\phi_i\) are, respectively, amplitudes,
periods and phases of the periodic effects.}. These terms come from
the FFT and correspond to the following main spectral lines of the
YS effect \cite{2002P&SS...50.1067L}:
\(\dot{\lambda}+\dot{\omega}\) (\(\simeq\)~257~days),
\(\dot{\lambda}-\dot{\omega}\) (\(\simeq\)~624~days),
\(\dot{\Omega}+\dot{\lambda}+\dot{\omega}\) (\(\simeq\)~485~days)
and \(\dot{\Omega}-\dot{\lambda}+\dot{\omega}\)
(\(\simeq\)~312~days) \footnote{These frequencies have been taken
fixed, while amplitudes and phases of the periodic terms
have been adjusted.}. For the linear trend slope we obtained a best
value of \(\Delta\dot{\omega}_{II}^{meas} = 3306.58\)~mas/yr, that
corresponds to a fractional discrepancy of about 0.03\% with respect
to the prediction of GR.
The result of the fit mainly depends on the time span of the data
analysis and on the number of periodic effects which are fitted
together with the linear term. The worst result we obtained, by
changing some of the initial conditions and the number of adjusted
parameters, has been a 0.21\% discrepancy between the value of the
recovered slope and the predicted one, see Table \ref{table:2}.
\begin{table}[h]
\begin{center}
\caption{{The fit results are mainly sensitive to the value of the
intercept \(a\) of the fitting function. For \(a\) fixed to the
value obtained from the residuals (280.9 mas), we obtained our best-fit
result independently of the initial fixed value for the other
parameters and the number of the periodic effects we added to the
main four of the YS effect. Conversely, by varying
the intercept up to \(\pm 50\) mas, value that corresponds to the
rms of the range residuals of the adjusted state vector,
we obtained our worst result. \(\Delta a\) is the
variation for the intercept while \(\delta a\) and \(\delta b\)
represent the adjustment of the given parameter}.} \label{table:2}
\newcommand{\m}{\hphantom{$-$}}
\newcommand{\cc}[1]{\multicolumn{1}{c}{#1}}
\renewcommand{\tabcolsep}{2pc} 
\renewcommand{\arraystretch}{1.2} 
\begin{tabular}{@{}ccc}
\hline
\(\Delta a \) [mas] &  \(\delta a \) [mas] &  \(\delta b/\Delta\dot{\omega}_{II}^{rel}\) [\%] \\
\hline
0 & 0  & + 0.03\\
\(\pm 10\) & -9.9  & + 0.06\\
\(\pm 20\) & -19.9  & + 0.10\\
\(\pm 30\) & -29.9  & + 0.13\\
\(\pm 40\) & -39.9  & + 0.16\\
\(\pm 50\) & -49.9  & + 0.21\\
\hline
\end{tabular}\\[2pt]
\end{center}
\end{table}
Therefore, for our analysis of the LAGEOS II pericenter
general relativistic advance we assume the following conservative
result:
\begin{equation}
\label{epsilon} \epsilon_{\omega}=1+(0.28\pm2.14)\cdot10^{-3}.
\end{equation}
The parameter \(\epsilon_{\omega}\) may be considered at the
post--Newtonian level, and measures possible deviations from
Einstein's GR, where \(\epsilon_{\omega}= 1\). In the case of
LAGEOS, we obtained a worst best fit, with at least a 1\%
discrepancy with respect to the prediction of GR
(\(\Delta\dot{\omega}_{I}^{rel} \simeq 3304.78\) mas/yr), due to the
smaller eccentricity and the larger perturbations produced by the
unmodeled YS effect \footnote{The analysis with EGM96 shows a smaller accuracy in
\(\epsilon_{\omega}\). The best result has a 2\%
discrepancy with respect to the GR prediction in the case of LAGEOS II.}.

The result obtained with the current analysis represents, to our
knowledge, the most accurate measurement for the pericenter advance
of a satellite orbiting the Earth ever made. In the PPN framework, it can be
considered as a 0.03\% measurement of the combination of the
\(\gamma\) and \(\beta\) parameters. Indeed, since
the leading contribution comes from Einstein's secular precession, we can consider
\(\epsilon_{\omega}\simeq\epsilon_E=(2+2\gamma-\beta)/3\)
\footnote{The total precession in the PPN formalism may be written as:
\(\Delta\dot{\omega}^{rel}\simeq\epsilon_E
\Delta\dot{\omega}^{E}+\epsilon_{LT}
\Delta\dot{\omega}^{LT}+\epsilon_{dS} \Delta\dot{\omega}^{dS}
+\ldots\), where the LT and de Sitter parameters are function of
\(\gamma\) only.}. The impact on the argument of pericenter of a
possible NLRI described via a Yukawa--like interaction has been
evaluated in \cite{2003PhLA..318..234L}, where also the contribution
of the main systematic effects has been estimated. The secular
effect is given by:
\begin{equation}
\Delta\dot{\omega}^{Yuk}\simeq8.2923586\cdot10^{11}\alpha \
\text{[mas/yr]} \label{yuk1}
\end{equation}
and it corresponds to the peak value at a range
\(\lambda = 6,081\)~km, very close to 1 Earth radius. Hence, we can
consider our measurement as an upper bound for
the strength of a possible long--range interaction; we obtain:
\begin{equation}
\label{yuk2} |\alpha|\simeq |(1.0 \pm 8.9)|\cdot 10^{-12}.
\end{equation}
This result represents a huge improvement in the constraint of the
strength \(\alpha\) at 1 Earth radius. Previous results using
Earth--LAGEOS and Lunar--LAGEOS measurements of \(GM\) where
confined at the level of \(10^{-5}\) and \(10^{-8}\). Our result for
\(\alpha\) is comparable with those obtained with Lunar Laser
Ranging (LLR) measurements at a characteristic scale of about 60
Earth radii, see Ref. \cite{2008ASSL..349..457M}.
With regard to the impact of the systematic errors on the
measurement performed so far, these are mainly related with the
uncertainty of the first even zonal harmonic \(J_2\)
\cite{2003PhLA..318..234L}, and we can preliminarily assume a 2\%
value. In a forthcoming paper a full characterization of the
systematic errors in the pericenter rate will be given, together
with the results we obtained for the two LAGEOS satellites' ascending
node longitude and inclination precessions.

%



\begin{acknowledgments}
The authors acknowledge the ILRS for providing high--quality laser
ranging data of the two LAGEOS satellites and two anonymous referees
for helpful comments. This work has been in part supported by the Italian Space Agency.
\end{acknowledgments}

\bibliography{Lucchesi-Peron_PRL_105_231103_arXiv}

\begin{thebibliography}{39}%
\makeatletter
\providecommand \@ifxundefined [1]{%
 \@ifx{#1\undefined}
}%
\providecommand \@ifnum [1]{%
 \ifnum #1\expandafter \@firstoftwo
 \else \expandafter \@secondoftwo
 \fi
}%
\providecommand \@ifx [1]{%
 \ifx #1\expandafter \@firstoftwo
 \else \expandafter \@secondoftwo
 \fi
}%
\providecommand \natexlab [1]{#1}%
\providecommand \enquote  [1]{``#1''}%
\providecommand \bibnamefont  [1]{#1}%
\providecommand \bibfnamefont [1]{#1}%
\providecommand \citenamefont [1]{#1}%
\providecommand \href@noop [0]{\@secondoftwo}%
\providecommand \href [0]{\begingroup \@sanitize@url \@href}%
\providecommand \@href[1]{\@@startlink{#1}\@@href}%
\providecommand \@@href[1]{\endgroup#1\@@endlink}%
\providecommand \@sanitize@url [0]{\catcode `\\12\catcode `\$12\catcode
  `\&12\catcode `\#12\catcode `\^12\catcode `\_12\catcode `\%12\relax}%
\providecommand \@@startlink[1]{}%
\providecommand \@@endlink[0]{}%
\providecommand \url  [0]{\begingroup\@sanitize@url \@url }%
\providecommand \@url [1]{\endgroup\@href {#1}{\urlprefix }}%
\providecommand \urlprefix  [0]{URL }%
\providecommand \Eprint [0]{\href }%
\@ifxundefined \urlstyle {%
  \providecommand \doi  [0]{\begingroup \@sanitize@url \@doi}%
  \providecommand \@doi [1]{\endgroup \@@startlink {\doibase
  #1}doi:\discretionary {}{}{}#1\@@endlink }%
}{%
  \providecommand \doi  [0]{doi:\discretionary{}{}{}\begingroup
  \urlstyle{rm}\Url }%
}%
\providecommand \doibase [0]{http://dx.doi.org/}%
\providecommand \Doi [0]{\begingroup \@sanitize@url \@Doi }%
\providecommand \@Doi  [1]{\endgroup\@@startlink{\doibase#1}\@@Doi}%
\providecommand \@@Doi [1]{#1\@@endlink}%
\providecommand \selectlanguage [0]{\@gobble}%
\providecommand \bibinfo  [0]{\@secondoftwo}%
\providecommand \bibfield  [0]{\@secondoftwo}%
\providecommand \translation [1]{[#1]}%
\providecommand \BibitemOpen [0]{}%
\providecommand \bibitemStop [0]{}%
\providecommand \bibitemNoStop [0]{.\EOS\space}%
\providecommand \EOS [0]{\spacefactor3000\relax}%
\providecommand \BibitemShut  [1]{\csname bibitem#1\endcsname}%
\bibitem [{\citenamefont {{Damour}}\ \emph {et~al.}(2002)\citenamefont
  {{Damour}}, \citenamefont {{Piazza}},\ and\ \citenamefont
  {{Veneziano}}}]{2002PhRvD..66d6007D}%
  \BibitemOpen
  \bibfield  {author} {\bibinfo {author} {\bibfnamefont {T.}~\bibnamefont
  {{Damour}}}, \bibinfo {author} {\bibfnamefont {F.}~\bibnamefont {{Piazza}}},
  \ and\ \bibinfo {author} {\bibfnamefont {G.}~\bibnamefont {{Veneziano}}},\
  }\Doi {10.1103/PhysRevD.66.046007} {\bibfield  {journal} {\bibinfo  {journal}
  {Phys. Rev. D},\ }\textbf {\bibinfo {volume} {66}},\ \bibinfo {pages}
  {046007} (\bibinfo {year} {2002})}\BibitemShut {NoStop}%
\bibitem [{\citenamefont {{Fischbach}}\ and\ \citenamefont {{et
  al.}}(1986)}]{1986PhRvL..56....3F}%
  \BibitemOpen
  \bibfield  {author} {\bibinfo {author} {\bibfnamefont {E.}~\bibnamefont
  {{Fischbach}}}\ and\ \bibinfo {author} {\bibnamefont {{et al.}}},\ }\Doi
  {10.1103/PhysRevLett.56.3} {\bibfield  {journal} {\bibinfo  {journal} {Phys.
  Rev. Lett.},\ }\textbf {\bibinfo {volume} {56}},\ \bibinfo {pages} {3}
  (\bibinfo {year} {1986})}\BibitemShut {NoStop}%
\bibitem [{\citenamefont {{Nordtvedt}}(1998)}]{1998Nordtvedt}%
  \BibitemOpen
  \bibfield  {author} {\bibinfo {author} {\bibfnamefont {K.}~\bibnamefont
  {{Nordtvedt}}},\ }\enquote {\bibinfo {title} {{LARES ASI Phase A Report}},}\
  \ (\bibinfo {year} {1998})\ pp.\ \bibinfo {pages} {34--37}\BibitemShut
  {NoStop}%
\bibitem [{\citenamefont {{Nordtvedt}}(2000)}]{2000PhRvD..61l2001N}%
  \BibitemOpen
  \bibfield  {author} {\bibinfo {author} {\bibfnamefont {K.}~\bibnamefont
  {{Nordtvedt}}},\ }\Doi {10.1103/PhysRevD.61.122001} {\bibfield  {journal}
  {\bibinfo  {journal} {Phys. Rev. D},\ }\textbf {\bibinfo {volume} {61}},\
  \bibinfo {pages} {122001} (\bibinfo {year} {2000})}\BibitemShut {NoStop}%
\bibitem [{Note1()}]{Note1}%
  \BibitemOpen
  \bibinfo {note} {Masses produce a gravitoelectric field, analogous to the
  electric field produced by charges, while mass currents produce a
  gravitomagnetic field, analogous to the magnetic field produced by electric
  currents.}\BibitemShut {Stop}%
\bibitem [{\citenamefont {{Einstein}}(1916)}]{1916AnP...354..769E}%
  \BibitemOpen
  \bibfield  {author} {\bibinfo {author} {\bibfnamefont {A.}~\bibnamefont
  {{Einstein}}},\ }\Doi {10.1002/andp.19163540702} {\bibfield  {journal}
  {\bibinfo  {journal} {Ann. Phys. (Leipzig)},\ }\textbf {\bibinfo {volume}
  {354}},\ \bibinfo {pages} {769} (\bibinfo {year} {1916})}\BibitemShut
  {NoStop}%
\bibitem [{\citenamefont {{de Sitter}}(1916)}]{1916MNRAS..77..155D}%
  \BibitemOpen
  \bibfield  {author} {\bibinfo {author} {\bibfnamefont {W.}~\bibnamefont {{de
  Sitter}}},\ }\href@noop {} {\bibfield  {journal} {\bibinfo  {journal} {Mon.
  Not. R. Astron. Soc.},\ }\textbf {\bibinfo {volume} {77}},\ \bibinfo {pages}
  {155} (\bibinfo {year} {1916})}\BibitemShut {NoStop}%
\bibitem [{\citenamefont {{Lense}}\ and\ \citenamefont
  {{Thirring}}(1918)}]{1918Lense-Thirring}%
  \BibitemOpen
  \bibfield  {author} {\bibinfo {author} {\bibfnamefont {J.}~\bibnamefont
  {{Lense}}}\ and\ \bibinfo {author} {\bibfnamefont {H.}~\bibnamefont
  {{Thirring}}},\ }\href@noop {} {\bibfield  {journal} {\bibinfo  {journal}
  {Phys. Z.},\ }\textbf {\bibinfo {volume} {19}},\ \bibinfo {pages} {156}
  (\bibinfo {year} {1918})}\BibitemShut {NoStop}%
\bibitem [{\citenamefont {{Mashhoon}}\ \emph {et~al.}(1984)\citenamefont
  {{Mashhoon}}, \citenamefont {{Hehl}},\ and\ \citenamefont
  {{Theiss}}}]{1984GReGr..16..711M}%
  \BibitemOpen
  \bibfield  {author} {\bibinfo {author} {\bibfnamefont {B.}~\bibnamefont
  {{Mashhoon}}}, \bibinfo {author} {\bibfnamefont {F.~W.}\ \bibnamefont
  {{Hehl}}}, \ and\ \bibinfo {author} {\bibfnamefont {D.~S.}\ \bibnamefont
  {{Theiss}}},\ }\Doi {10.1007/BF00762913} {\bibfield  {journal} {\bibinfo
  {journal} {Gen. Rel. Grav.},\ }\textbf {\bibinfo {volume} {16}},\ \bibinfo
  {pages} {711} (\bibinfo {year} {1984})}\BibitemShut {NoStop}%
\bibitem [{\citenamefont {{Ciufolini}}\ and\ \citenamefont
  {{Wheeler}}(1995)}]{1995grin.book.....C}%
  \BibitemOpen
  \bibfield  {author} {\bibinfo {author} {\bibfnamefont {I.}~\bibnamefont
  {{Ciufolini}}}\ and\ \bibinfo {author} {\bibfnamefont {J.~A.}\ \bibnamefont
  {{Wheeler}}},\ }\href@noop {} {\emph {\bibinfo {title} {{Gravitation and
  inertia}}}}\ (\bibinfo  {publisher} {Princeton University Press},\ \bibinfo
  {address} {Princeton},\ \bibinfo {year} {1995})\BibitemShut {NoStop}%
\bibitem [{\citenamefont {{Ciufolini}}(2010)}]{2010NewA...15..332C}%
  \BibitemOpen
  \bibfield  {author} {\bibinfo {author} {\bibfnamefont {I.}~\bibnamefont
  {{Ciufolini}}},\ }\Doi {10.1016/j.newast.2009.08.004} {\bibfield  {journal}
  {\bibinfo  {journal} {New Astron.},\ }\textbf {\bibinfo {volume} {15}},\
  \bibinfo {pages} {332} (\bibinfo {year} {2010})}\BibitemShut {NoStop}%
\bibitem [{\citenamefont {{Rubincam}}(1977)}]{1977CeMec..15...21R}%
  \BibitemOpen
  \bibfield  {author} {\bibinfo {author} {\bibfnamefont {D.~P.}\ \bibnamefont
  {{Rubincam}}},\ }\Doi {10.1007/BF01229045} {\bibfield  {journal} {\bibinfo
  {journal} {Celest. Mech.},\ }\textbf {\bibinfo {volume} {15}},\ \bibinfo
  {pages} {21} (\bibinfo {year} {1977})}\BibitemShut {NoStop}%
\bibitem [{\citenamefont {{Pavlis}}\ and\ \citenamefont {{et
  al.}}(1998)}]{1998pavlis}%
  \BibitemOpen
  \bibfield  {author} {\bibinfo {author} {\bibfnamefont {D.~E.}\ \bibnamefont
  {{Pavlis}}}\ and\ \bibinfo {author} {\bibnamefont {{et al.}}},\ }\href@noop
  {} {\emph {\bibinfo {title} {{GEODYN II Operations Manual}}}},\ \bibinfo
  {organization} {NASA GSFC} (\bibinfo {year} {1998})\BibitemShut {NoStop}%
\bibitem [{\citenamefont {{Ashby}}\ and\ \citenamefont
  {{Bertotti}}(1984)}]{1984PhRvL..52..485A}%
  \BibitemOpen
  \bibfield  {author} {\bibinfo {author} {\bibfnamefont {N.}~\bibnamefont
  {{Ashby}}}\ and\ \bibinfo {author} {\bibfnamefont {B.}~\bibnamefont
  {{Bertotti}}},\ }\href@noop {} {\bibfield  {journal} {\bibinfo  {journal}
  {Phys. Rev. Lett.},\ }\textbf {\bibinfo {volume} {52}},\ \bibinfo {pages}
  {485} (\bibinfo {year} {1984})}\BibitemShut {NoStop}%
\bibitem [{\citenamefont {{Huang}}\ \emph {et~al.}(1990)\citenamefont
  {{Huang}}, \citenamefont {{Ries}}, \citenamefont {{Tapley}},\ and\
  \citenamefont {{Watkins}}}]{1990CeMDA..48..167H}%
  \BibitemOpen
  \bibfield  {author} {\bibinfo {author} {\bibfnamefont {C.}~\bibnamefont
  {{Huang}}}, \bibinfo {author} {\bibfnamefont {J.~C.}\ \bibnamefont {{Ries}}},
  \bibinfo {author} {\bibfnamefont {B.~D.}\ \bibnamefont {{Tapley}}}, \ and\
  \bibinfo {author} {\bibfnamefont {M.~M.}\ \bibnamefont {{Watkins}}},\ }\Doi
  {10.1007/BF00049512} {\bibfield  {journal} {\bibinfo  {journal} {Celest.
  Mech. Dyn. Astron.},\ }\textbf {\bibinfo {volume} {48}},\ \bibinfo {pages}
  {167} (\bibinfo {year} {1990})}\BibitemShut {NoStop}%
\bibitem [{\citenamefont {{Andr\`{e}s de la Fuente}}(2007)}]{2007Andres}%
  \BibitemOpen
  \bibfield  {author} {\bibinfo {author} {\bibfnamefont {J.~I.}\ \bibnamefont
  {{Andr\`{e}s de la Fuente}}},\ }\href@noop {} {Ph.D. thesis},\ \bibinfo
  {school} {Delft University Press} (\bibinfo {year} {2007})\BibitemShut
  {NoStop}%
\bibitem [{\citenamefont {{Milani}}\ \emph {et~al.}(1987)\citenamefont
  {{Milani}}, \citenamefont {{Nobili}},\ and\ \citenamefont
  {{Farinella}}}]{1987nongrav.book.....M}%
  \BibitemOpen
  \bibfield  {author} {\bibinfo {author} {\bibfnamefont {A.}~\bibnamefont
  {{Milani}}}, \bibinfo {author} {\bibfnamefont {A.~M.}\ \bibnamefont
  {{Nobili}}}, \ and\ \bibinfo {author} {\bibfnamefont {P.}~\bibnamefont
  {{Farinella}}},\ }\href@noop {} {\emph {\bibinfo {title} {{Non-gravitational
  perturbations and satellite geodesy}}}}\ (\bibinfo  {publisher} {Adam
  Hilger},\ \bibinfo {address} {Bristol},\ \bibinfo {year} {1987})\BibitemShut
  {NoStop}%
\bibitem [{Note2()}]{Note2}%
  \BibitemOpen
  \bibinfo {note} {LAGEOS has an almost circular orbit, with an eccentricity
  \(e_{I} \simeq 0.004\), a semimajor axis \(a_{I} \simeq 12,270\)~km and an
  inclination over the Earth's equator \(i_{I} \simeq 109.8^{\circ }\). LAGEOS
  II corresponding elements are: \(e_{II} \simeq 0.014\), \(a_{II} \simeq
  12,162\)~km and \(i_{II} \simeq 52.66^{\circ }\).}\BibitemShut {Stop}%
\bibitem [{\citenamefont {{Pearlman}}\ \emph {et~al.}(2002)\citenamefont
  {{Pearlman}}, \citenamefont {{Degnan}},\ and\ \citenamefont
  {{Bosworth}}}]{2002AdSpR..30..135P}%
  \BibitemOpen
  \bibfield  {author} {\bibinfo {author} {\bibfnamefont {M.~R.}\ \bibnamefont
  {{Pearlman}}}, \bibinfo {author} {\bibfnamefont {J.~J.}\ \bibnamefont
  {{Degnan}}}, \ and\ \bibinfo {author} {\bibfnamefont {J.~M.}\ \bibnamefont
  {{Bosworth}}},\ }\href@noop {} {\bibfield  {journal} {\bibinfo  {journal}
  {Adv. Space Res.},\ }\textbf {\bibinfo {volume} {30}},\ \bibinfo {pages}
  {135} (\bibinfo {year} {2002})}\BibitemShut {NoStop}%
\bibitem [{\citenamefont {{McCarthy}}\ and\ \citenamefont
  {{Petit}}(2004)}]{2004IERS-Conv-2003}%
  \BibitemOpen
  \bibfield  {author} {\bibinfo {author} {\bibfnamefont {D.~D.}\ \bibnamefont
  {{McCarthy}}}\ and\ \bibinfo {author} {\bibfnamefont {G.}~\bibnamefont
  {{Petit}}},\ }\href@noop {} {\emph {\bibinfo {title} {IERS Conventions
  (2003)}}},\ \bibinfo {type} {IERS Technical Note}\ \bibinfo {number} {32}\
  (\bibinfo  {institution} {IERS},\ \bibinfo {year} {2004})\BibitemShut
  {NoStop}%
\bibitem [{\citenamefont {{Nordtvedt}}(1968)}]{1968PhRv..169.1017N}%
  \BibitemOpen
  \bibfield  {author} {\bibinfo {author} {\bibfnamefont {K.}~\bibnamefont
  {{Nordtvedt}}},\ }\Doi {10.1103/PhysRev.169.1017} {\bibfield  {journal}
  {\bibinfo  {journal} {Phys. Rev.},\ }\textbf {\bibinfo {volume} {169}},\
  \bibinfo {pages} {1017} (\bibinfo {year} {1968})}\BibitemShut {NoStop}%
\bibitem [{\citenamefont {{Will}}(1971)}]{1971ApJ...163..611W}%
  \BibitemOpen
  \bibfield  {author} {\bibinfo {author} {\bibfnamefont {C.~M.}\ \bibnamefont
  {{Will}}},\ }\Doi {10.1086/150804} {\bibfield  {journal} {\bibinfo  {journal}
  {Astrophys. J.},\ }\textbf {\bibinfo {volume} {163}},\ \bibinfo {pages} {611}
  (\bibinfo {year} {1971})}\BibitemShut {NoStop}%
\bibitem [{Note3()}]{Note3}%
  \BibitemOpen
  \bibinfo {note} {It is necessary to decompose the long arc in a number of
  shorter arcs (15 days in our analysis), not causally connected, and solve for
  the satellite state vector for each arc together with a set of parameters in
  order to absorb unmodeled or poorly modeled perturbations over the
  arc.}\BibitemShut {Stop}%
\bibitem [{\citenamefont {{Lucchesi}}\ and\ \citenamefont
  {{Balmino}}(2006)}]{2006P&SS...54..581L}%
  \BibitemOpen
  \bibfield  {author} {\bibinfo {author} {\bibfnamefont {D.~M.}\ \bibnamefont
  {{Lucchesi}}}\ and\ \bibinfo {author} {\bibfnamefont {G.}~\bibnamefont
  {{Balmino}}},\ }\Doi {10.1016/j.pss.2006.03.001} {\bibfield  {journal}
  {\bibinfo  {journal} {Plan. Space Sci.},\ }\textbf {\bibinfo {volume} {54}},\
  \bibinfo {pages} {581} (\bibinfo {year} {2006})}\BibitemShut {NoStop}%
\bibitem [{\citenamefont {{Ciufolini}}\ \emph {et~al.}(1996)\citenamefont
  {{Ciufolini}}, \citenamefont {{Lucchesi}}, \citenamefont {{Vespe}},\ and\
  \citenamefont {{Mandiello}}}]{1996NCimA.109..575C}%
  \BibitemOpen
  \bibfield  {author} {\bibinfo {author} {\bibfnamefont {I.}~\bibnamefont
  {{Ciufolini}}}, \bibinfo {author} {\bibfnamefont {D.}~\bibnamefont
  {{Lucchesi}}}, \bibinfo {author} {\bibfnamefont {F.}~\bibnamefont {{Vespe}}},
  \ and\ \bibinfo {author} {\bibfnamefont {A.}~\bibnamefont {{Mandiello}}},\
  }\Doi {10.1007/BF02731140} {\bibfield  {journal} {\bibinfo  {journal} {Nuovo
  Cim. A},\ }\textbf {\bibinfo {volume} {109}},\ \bibinfo {pages} {575}
  (\bibinfo {year} {1996})}\BibitemShut {NoStop}%
\bibitem [{\citenamefont {{Ciufolini}}\ \emph {et~al.}(1997)\citenamefont
  {{Ciufolini}}, \citenamefont {{Lucchesi}}, \citenamefont {{Vespe}},\ and\
  \citenamefont {{Chieppa}}}]{1997EL.....39..359C}%
  \BibitemOpen
  \bibfield  {author} {\bibinfo {author} {\bibfnamefont {I.}~\bibnamefont
  {{Ciufolini}}}, \bibinfo {author} {\bibfnamefont {D.}~\bibnamefont
  {{Lucchesi}}}, \bibinfo {author} {\bibfnamefont {F.}~\bibnamefont {{Vespe}}},
  \ and\ \bibinfo {author} {\bibfnamefont {F.}~\bibnamefont {{Chieppa}}},\
  }\Doi {10.1209/epl/i1997-00362-7} {\bibfield  {journal} {\bibinfo  {journal}
  {Europhys. Lett.},\ }\textbf {\bibinfo {volume} {39}},\ \bibinfo {pages}
  {359} (\bibinfo {year} {1997})}\BibitemShut {NoStop}%
\bibitem [{\citenamefont {{Ciufolini}}\ \emph {et~al.}(1998)\citenamefont
  {{Ciufolini}}, \citenamefont {{Pavlis}}, \citenamefont {{Chieppa}},
  \citenamefont {{Fernandes-Vieira}},\ and\ \citenamefont
  {{Perez-Mercader}}}]{1998Sci...279.2100C}%
  \BibitemOpen
  \bibfield  {author} {\bibinfo {author} {\bibfnamefont {I.}~\bibnamefont
  {{Ciufolini}}}, \bibinfo {author} {\bibfnamefont {E.}~\bibnamefont
  {{Pavlis}}}, \bibinfo {author} {\bibfnamefont {F.}~\bibnamefont {{Chieppa}}},
  \bibinfo {author} {\bibfnamefont {E.}~\bibnamefont {{Fernandes-Vieira}}}, \
  and\ \bibinfo {author} {\bibfnamefont {J.}~\bibnamefont {{Perez-Mercader}}},\
  }\Doi {10.1126/science.279.5359.2100} {\bibfield  {journal} {\bibinfo
  {journal} {Science},\ }\textbf {\bibinfo {volume} {279}},\ \bibinfo {pages}
  {2100} (\bibinfo {year} {1998})}\BibitemShut {NoStop}%
\bibitem [{\citenamefont {{Ciufolini}}\ and\ \citenamefont
  {{Pavlis}}(2004)}]{2004Natur.431..958C}%
  \BibitemOpen
  \bibfield  {author} {\bibinfo {author} {\bibfnamefont {I.}~\bibnamefont
  {{Ciufolini}}}\ and\ \bibinfo {author} {\bibfnamefont {E.~C.}\ \bibnamefont
  {{Pavlis}}},\ }\Doi {10.1038/nature03007} {\bibfield  {journal} {\bibinfo
  {journal} {Nature},\ }\textbf {\bibinfo {volume} {431}},\ \bibinfo {pages}
  {958} (\bibinfo {year} {2004})}\BibitemShut {NoStop}%
\bibitem [{\citenamefont {{Ciufolini}}\ \emph {et~al.}(2006)\citenamefont
  {{Ciufolini}}, \citenamefont {{Pavlis}},\ and\ \citenamefont
  {{Peron}}}]{2006NewA...11..527C}%
  \BibitemOpen
  \bibfield  {author} {\bibinfo {author} {\bibfnamefont {I.}~\bibnamefont
  {{Ciufolini}}}, \bibinfo {author} {\bibfnamefont {E.~C.}\ \bibnamefont
  {{Pavlis}}}, \ and\ \bibinfo {author} {\bibfnamefont {R.}~\bibnamefont
  {{Peron}}},\ }\Doi {10.1016/j.newast.2006.02.001} {\bibfield  {journal}
  {\bibinfo  {journal} {New Astron.},\ }\textbf {\bibinfo {volume} {11}},\
  \bibinfo {pages} {527} (\bibinfo {year} {2006})}\BibitemShut {NoStop}%
\bibitem [{\citenamefont {{Lemoine}}\ and\ \citenamefont {{et
  al.}}(1998)}]{1998Lemoine}%
  \BibitemOpen
  \bibfield  {author} {\bibinfo {author} {\bibfnamefont {F.~G.}\ \bibnamefont
  {{Lemoine}}}\ and\ \bibinfo {author} {\bibnamefont {{et al.}}},\ }\href@noop
  {} {}\bibinfo {type} {Technical Paper}\ \bibinfo {number}
  {NASA/TP--1998--206861}\ (\bibinfo {year} {1998})\BibitemShut {NoStop}%
\bibitem [{Note4()}]{Note4}%
  \BibitemOpen
  \bibinfo {note} {In this way, the gravity model has more or less the same
  sensitivity to the low and the high degree terms, especially because of the
  different altitudes of the satellites. EGM96 is characterized by uncalibrated
  formal errors with a relative high correlation between the
  coefficients.}\BibitemShut {Stop}%
\bibitem [{\citenamefont {{Reigber}}\ and\ \citenamefont {{et
  al.}}(2005)}]{2005JGeo...39....1R}%
  \BibitemOpen
  \bibfield  {author} {\bibinfo {author} {\bibfnamefont {C.}~\bibnamefont
  {{Reigber}}}\ and\ \bibinfo {author} {\bibnamefont {{et al.}}},\ }\Doi
  {10.1016/j.jog.2004.07.001} {\bibfield  {journal} {\bibinfo  {journal} {J.
  Geodyn.},\ }\textbf {\bibinfo {volume} {39}},\ \bibinfo {pages} {1} (\bibinfo
  {year} {2005})}\BibitemShut {NoStop}%
\bibitem [{Note5()}]{Note5}%
  \BibitemOpen
  \bibinfo {note} {That is, \(\Delta \omega _{fit}=a+bt+\DOTSB \sum@ \slimits@
  _i C_i\protect \qopname \relax o{sin}(2\pi t/T_i+\phi _i)\), where
  \(b=\protect \mathaccentV {dot}05F{\omega }_{II}^{meas}\) while \(C_i\),
  \(T_i\) and \(\phi _i\) are, respectively, amplitudes, periods and phases of
  the periodic effects.}\BibitemShut {Stop}%
\bibitem [{\citenamefont {{Lucchesi}}(2002)}]{2002P&SS...50.1067L}%
  \BibitemOpen
  \bibfield  {author} {\bibinfo {author} {\bibfnamefont {D.~M.}\ \bibnamefont
  {{Lucchesi}}},\ }\href@noop {} {\bibfield  {journal} {\bibinfo  {journal}
  {Plan. Space Sci.},\ }\textbf {\bibinfo {volume} {50}},\ \bibinfo {pages}
  {1067} (\bibinfo {year} {2002})}\BibitemShut {NoStop}%
\bibitem [{Note6()}]{Note6}%
  \BibitemOpen
  \bibinfo {note} {These frequencies have been taken fixed, while amplitudes
  and phases of the periodic terms have been adjusted.}\BibitemShut {Stop}%
\bibitem [{Note7()}]{Note7}%
  \BibitemOpen
  \bibinfo {note} {The analysis with EGM96 shows a smaller accuracy in
  \(\epsilon _{\omega }\). The best result has a 2\% discrepancy with respect
  to the GR prediction in the case of LAGEOS II.}\BibitemShut {Stop}%
\bibitem [{Note8()}]{Note8}%
  \BibitemOpen
  \bibinfo {note} {The total precession in the PPN formalism may be written as:
  \(\Delta \protect \mathaccentV {dot}05F{\omega }^{rel}\simeq \epsilon _E
  \Delta \protect \mathaccentV {dot}05F{\omega }^{E}+\epsilon _{LT} \Delta
  \protect \mathaccentV {dot}05F{\omega }^{LT}+\epsilon _{dS} \Delta \protect
  \mathaccentV {dot}05F{\omega }^{dS} +\protect \ldots \), where the LT and de
  Sitter parameters are function of \(\gamma \) only.}\BibitemShut {Stop}%
\bibitem [{\citenamefont {{Lucchesi}}(2003)}]{2003PhLA..318..234L}%
  \BibitemOpen
  \bibfield  {author} {\bibinfo {author} {\bibfnamefont {D.~M.}\ \bibnamefont
  {{Lucchesi}}},\ }\Doi {10.1016/S0375-9601(03)01213-1} {\bibfield  {journal}
  {\bibinfo  {journal} {Phys. Lett. A},\ }\textbf {\bibinfo {volume} {318}},\
  \bibinfo {pages} {234} (\bibinfo {year} {2003})}\BibitemShut {NoStop}%
\bibitem [{\citenamefont {{M{\"u}ller}}\ \emph {et~al.}(2008)\citenamefont
  {{M{\"u}ller}}, \citenamefont {{Williams}},\ and\ \citenamefont
  {{Turyshev}}}]{2008ASSL..349..457M}%
  \BibitemOpen
  \bibfield  {author} {\bibinfo {author} {\bibfnamefont {J.}~\bibnamefont
  {{M{\"u}ller}}}, \bibinfo {author} {\bibfnamefont {J.~G.}\ \bibnamefont
  {{Williams}}}, \ and\ \bibinfo {author} {\bibfnamefont {S.~G.}\ \bibnamefont
  {{Turyshev}}},\ }in\ \Doi {10.1007/978-3-540-34377-6_21} {\emph {\bibinfo
  {booktitle} {{Lasers, Clocks and Drag-Free Control: Exploration of
  Relativistic Gravity in Space}}}},\ \bibinfo {editor} {edited by\ \bibinfo
  {editor} {\bibnamefont {{H.~Dittus, C.~Lammerzahl, \& S.~G.~Turyshev}}}}\
  (\bibinfo {year} {2008})\ p.\ \bibinfo {pages} {457}\BibitemShut {NoStop}%
\end{thebibliography}%

\end{document}